\documentclass[a4paper,12pt]{article}

\usepackage{psfrag}
\usepackage{mathrsfs}

\usepackage{graphicx}
\usepackage[margin=15pt,font=small,labelfont=bf,labelsep=period]{caption}
\usepackage{amsmath}
\usepackage{amssymb}
\usepackage{amsfonts}
\usepackage{mathtools}
\usepackage{bm}
\usepackage{dsfont}
\usepackage{cite}
\usepackage{setspace}
\usepackage{environ,letltxmacro}
\usepackage{tikz}
\usetikzlibrary{arrows,decorations.pathreplacing}
\usepackage{enumitem}
\usepackage{hyperref}
\usepackage{relsize}
\usepackage[a4paper,textwidth=16.04cm,textheight=22cm,footskip=15mm]{geometry}

\usepackage{empheq}
\usepackage{environ}
\NewEnviron{boxalign}{\begin{empheq}[box=\fbox]{align} \BODY \end{empheq}}
\usepackage{color}

\numberwithin{equation}{section}

\LetLtxMacro\oldequation\equation
\LetLtxMacro\endoldequation\endequation
\let\equation\relax
\let\endequation\relax
\NewEnviron{equation}[1][\empty]{%
 \ifx \empty#1
  \oldequation \BODY \endoldequation
 \else
  \ifx b#1
   \oldequation \fbox{$\displaystyle \BODY $} \endoldequation
  \else
   \oldequation \BODY \endoldequation
  \fi
 \fi}

\begin{document}

\begin{titlepage}

\title{
\begin{flushright}
\normalsize{MITP/18-026}
\bigskip
\vspace{1cm}
\end{flushright}
Finite Entanglement Entropy in Asymptotically Safe Quantum Gravity \\[2mm]
}

\date{}

\author{Carlo Pagani and Martin Reuter\\[3mm]
{\small Institute of Physics, 
Johannes Gutenberg University Mainz,}\\[-0.2em]
{\small Staudingerweg 7, D--55099 Mainz, Germany}
}

\maketitle
\thispagestyle{empty}

\vspace{2mm}
\begin{abstract}

Entanglement entropies  calculated in the framework of quantum field theory on classical,
flat or curved, spacetimes are known to show an intriguing area law in four dimensions,
but they are also notorious for their quadratic ultraviolet divergences.
In this paper we demonstrate that the analogous entanglement entropies when computed
within the Asymptotic Safety approach to background independent quantum gravity are
perfectly free from such divergences. We argue that the divergences are an artifact due to the
over-idealization of a rigid, classical spacetime geometry which is insensitive to the quantum dynamics.

\end{abstract}

\end{titlepage}

\newpage

\begin{spacing}{1.1}


\section{Introduction}

One of the most remarkable, and in a way enigmatic, properties of
Quantum Mechanics is the occurrence of entangled states and the possibility
that local measurements instantaneously affect the result of local
measurements far away. While deeply intriguing as a physical phenomenon
in its own right, entanglement also received considerable attention
from an ``applied'' perspective, being at the heart of many modern
developments in quantum computation and information theory for example.
An improved understanding of the entanglement structure of quantum
many body systems allowed in particular developing new numerical algorithms
which can help in lowering the computational effort of the simulations \cite{HarocheBook,BengtssonBook}.

A frequently used quantity that can quantify the amount of entanglement,
at least in pure quantum states, is the \emph{entanglement entropy}.
Let $\rho=|\psi\rangle\langle\psi|$ denote the density operator of
an arbitrary quantum system in the pure state $|\psi\rangle$. We
assume that the pertinent Hilbert space is a direct product ${\cal H}={\cal H}_{A}\otimes{\cal H}_{B}$,
and that we are only interested in predictions for measurements which
affect the subspace ${\cal H}_{A}$ alone. Such predictions are encoded
in the reduced density matrix $\rho_{A}=\mbox{Tr}_{B}\left[\rho\right]$,
where $\mbox{Tr}_{B}$ denotes the partial trace over ${\cal H}_{B}$.
Then, by definition, the entanglement entropy related to the $A$-$B$
decomposition of the total system equals the von Neumann entropy of
$\rho_{A}$:
\begin{eqnarray}
{\cal S}_{A} & = & -\mbox{Tr}\left[\rho_{A}\log\rho_{A}\right]\,.\label{eq:def-von-neumann-entropy-A-1}
\end{eqnarray}
In practical calculation ${\cal S}_{A}$ is often represented as
the limit
\begin{eqnarray}
{\cal S}_{A} & = & -\lim_{n\rightarrow1}\frac{\partial}{\partial n}\mbox{Tr}\left[\rho_{A}^{n}\right]\,,\label{eq:S_A-via-rho_A-to-n-2}
\end{eqnarray}
and the replica trick is invoked in order to evaluate $\mbox{Tr}\left[\rho_{A}^{n}\right]$.
The latter consists in calculating $\mbox{Tr}\left[\rho_{A}^{n}\right]$
for positive integers $n$ and then analytically continue it to a domain in the complex
plane. If this step can be justified, calculating $\mbox{Tr}\left[\rho_{A}^{n}\right]$
can be seen to boil down to the evaluation of a certain partition
function; it generalizes the ordinary partition function governing
the quantum system considered in that it is defined over a modified, more complicated
base space, a kind of Riemann surface which may cover the original base
space more than once.

Consider for example a free quantum field on Minkowski space, with
coordinates $\left(t,x,y,z\right)$, and introduce a surface $\Sigma$
by the condition $t=0,\,x=0$.
This surface separates the time slice at $t=0$
in two parts, $x<0$ and $x>0$. 
If we now define the subsystems $A$ and
$B$ as comprised of the field degrees of freedom at $x<0$ and $x>0$, respectively, the ensuing
entanglement entropy ${\cal S}_{A}\equiv{\cal S}$ is given by~\cite{Callan:1994py,Kabat:1994vj,Solodukhin:2011gn}
\begin{eqnarray}
{\cal S} & = & \left[1+2\pi\frac{d}{d\delta}\right]\log Z_{\delta}\Bigr|_{\delta=0}\,.\label{eq:S-via-Z-with-def-angle-delta-5}
\end{eqnarray}
Here $Z_{\delta}$ is a standard partition function, albeit not on a Minkowski space but on
on a conical spacetime with a deficit angle $\delta$. Evaluating (\ref{eq:S-via-Z-with-def-angle-delta-5}),
ultraviolet (UV) divergences are encountered, and a short distance
cutoff, $\varepsilon$, needs to be introduced, yielding
\begin{eqnarray}
{\cal S} & = & \frac{\nu}{48\pi}\frac{A}{\varepsilon^{2}}\,.\label{eq:EE-of-free-fields-10}
\end{eqnarray}
In this formula, $\nu$ is a constant which depends on the type of field ($\nu=1$
for a scalar, for example) and $A$ denotes the area of $\Sigma$.
Hence, it is meaningful to speak of an entropy per area when $\Sigma$
is infinitely extended. However, ${\cal S}/A$ suffers from an UV
divergence, it diverges quadratically when the cutoff is removed ($\varepsilon\rightarrow0$).

The result (\ref{eq:EE-of-free-fields-10}) is valid under more general
conditions actually. On any (non-dynamical) curved spacetime, and
for an arbitrary closed smooth surface $\Sigma$, equation (\ref{eq:EE-of-free-fields-10})
gives the leading order contribution to the entanglement entropy.

Clearly, the physical interpretation of (\ref{eq:EE-of-free-fields-10})
is hampered by its UV divergence which cannot be ``renormalized away''
straightforwardly. Nevertheless, ever since its first discovery \cite{Sorkin:1985bu,Bombelli:1986rw,Srednicki:1993im},
the fact that ${\cal S}$ is proportional to the surface area rather
than the volume of the subsystem traced over has sparked considerable
interest and research activities
\cite{Hawking:1974sw,Bekenstein:1973ur,Hawking:1976de,Wald:1999vt,Sorkin:1985bu,Bombelli:1986rw,Srednicki:1993im,Frolov:1993ym,
Susskind:1994sm,Jacobson:1994iw,Callan:1994py,Solodukhin:1994yz,Fursaev:1994pq,Demers:1995dq,Kabat:1995eq,Larsen:1995ax,
Jacobson:2012ek,Cooperman:2013iqr}. 
One of the reasons is clearly the
similarity of (\ref{eq:EE-of-free-fields-10}) and the Bekenstein-Hawking
entropy in black hole thermodynamics,
\begin{eqnarray}
{\cal S}_{{\rm BH}} & = & \frac{A}{4G}\,,\label{eq:BH-entropy-11}
\end{eqnarray}
with $A$ denoting the area of the horizon now \cite{Hawking:1974sw,Bekenstein:1973ur,Hawking:1976de}. 
This similarity inspired attempts to partially or fully explain ${\cal S}_{{\rm BH}}$ as an entanglement
entropy, and thereby absorb the divergence of ${\cal S}$ in
a renormalized Newton constant. (We refer to \cite{Solodukhin:2011gn}
for a comprehensive account.)

The present paper is dedicated to the entanglement entropy (\ref{eq:EE-of-free-fields-10})
in its own right, i.e. without reference to black holes or other special
systems. Trying to pin down the physical origin of its quadratic divergence,
we are going to analyze what happens to the entanglement entropy when the above setting of quantum field
theory on classical spacetimes is generalized to full-fledged \emph{background
independent quantum gravity }\cite{Ashtekar:2014kba}. Concretely, we shall employ
the Asymptotic Safety approach \cite{W80,R98} to Quantum Einstein Gravity
(QEG) \cite{NR06,Percacci:2007sz,RSBook,Reuter:2012id,Reuter:2012xf}.

As we are going to argue, the divergence present in the standard result (\ref{eq:EE-of-free-fields-10})
originates from the fact that it answers, or tries to answer, an \emph{unphysical
question} that could never arise in a real physical experiment. The
over-idealization consists in considering ``test fields'' on an externally
prescribed classical spacetime. Instead, if the entanglement is studied
in a universe where the geometry is free to adjust itself dynamically
according to the gravitational dynamics implied by Asymptotic Safety,
the corresponding entropy turns out to be \emph{finite}. 

This is even the more remarkable as a number of quantum gravity models are known to
fail in rendering the entropy finite \cite{Arzano:2017mdp}.
It should be also emphasized that the proposed non-perturbative mechanism for
achieving a finite entanglement entropy does not rely on ``hiding''
its divergences in Newton's constant or similar couplings which parametrize the action functional. 

The rest of this paper is organized as follows. As a preparation we
briefly recall in section \ref{sec:Entanglement-entropy-on-rigid-background}
the derivation of equation (\ref{eq:EE-of-free-fields-10}) for classical
spacetimes. We also show how it relates in a natural way to the Effective
Average Action (EAA), the scale dependent functional that is 
used in the Asymptotic Safety program. Then, in section \ref{sec:Entanglement-entropy-in-QEG}, we proceed
to QEG and analyze the entanglement entropy in a universe with a scale dependent
spacetime geometry which is governed by an asymptotically safe renormalization
group flow.

\section{Entanglement entropy on a rigid background \label{sec:Entanglement-entropy-on-rigid-background}}

{\bf (A)} For any free matter field $\Phi$, governed by a quadratic
action $S\left[\Phi\right]$, the evaluation of the entropy by means of equation
(\ref{eq:S-via-Z-with-def-angle-delta-5}) consists in computing a
one-loop determinant on a locally flat spacetime with a conical defect,
$\log Z_{\delta}=-\frac{1}{2}\log\det\left(S^{\left(2\right)}\right)$.
Here $S^{\left(2\right)}$ denotes the Hessian operator of $S$. For
a scalar, say, $S^{\left(2\right)}=-\Box+m^{2}$. Standard manipulations
lead to the regularized proper time representation
\begin{eqnarray}
\log Z_{\delta} & = & \frac{1}{2}\int_{\varepsilon^{2}}^{\infty}\frac{dt}{t}K_{\delta}\left(t\right)\,,\,\mbox{with }K_{\delta}\left(t\right)\equiv\mbox{Tr}\left[e^{-tS^{\left(2\right)}}\right]\,.\label{eq:one-loop-EA-with-deficit-angle-20}
\end{eqnarray}
Here the length parameter $\varepsilon$ is introduced as a short distance
cutoff in order to cure the divergence of the $t$-integral at the
lower limit. So the essential ingredient we need is the heat kernel
$K_\delta\left(t\right)$ as a function of the deficit angle $\delta$,
\begin{eqnarray}
{\cal S} & = & \frac{1}{2}\lim_{\delta\rightarrow0}\int_{\varepsilon^{2}}^{\infty}\frac{dt}{t}\left[1+2\pi\frac{d}{d\delta}\right]K_{\delta}\left(t\right)\,.\label{eq:one-loop-S-via-HK-with-deficit-angle-21}
\end{eqnarray}
For a real, massless, minimally coupled scalar, the relevant part of
$K_\delta\left(t\right)$ can be found to be \cite{Dowker:1977zj,Fursaev:1994in}:
\begin{eqnarray}
K\left(t\right) & = & \frac{A}{\left(4\pi t\right)}\left[\frac{\pi L^{2}}{\left(4\pi t\right)}\left(1-\frac{\delta}{2\pi}\right)+\frac{\delta}{12\pi}+O\left(\delta^{2}\right)+O\left(\frac{t}{L^{2}}\right)\right]\,.\label{eq:explicit-HK-with-deficit-angle-22}
\end{eqnarray}
Here we set $AL^{2}\equiv \int d^{4}x$ for the $4D$ Euclidean volume.
Using (\ref{eq:explicit-HK-with-deficit-angle-22}) in (\ref{eq:one-loop-S-via-HK-with-deficit-angle-21})
one obtains exactly the anticipated result for the entanglement entropy,
equation (\ref{eq:EE-of-free-fields-10}), with $\nu=1$ for the real
scalar. Other systems of (higher spin) free fields lead to an analogous
formula with other values of the finite constant $\nu$, see \cite{Solodukhin:2011gn}
for a detailed discussion.\\
{\bf (B)} As a further preparation for the case of quantum gravity let us explain how the
above standard calculation should be interpreted within the general framework
of the EAA and the functional renormalization group \cite{W93}. 

The EAA for a scalar on a classical spacetime, $\Gamma_{k}\left[\Phi\right]$,
can be seen as the ordinary effective action for a field whose bare
action under the functional integral has been augmented by a mode cutoff
term: $S\left[\Phi\right]\rightarrow S\left[\Phi\right]+\frac{1}{2}\int\Phi{\cal R}_{k}\Phi$.
The operator ${\cal R}_{k}\equiv k^{2}R^{\left(0\right)}\left(-\Box/k^{2}\right)$
implements an infrared (IR) cutoff by giving a non-zero mass square
${\cal R}_{k}=k^{2}$ the low momentum modes contained in $\Phi$,
while annihilating the others, ${\cal R}_{k}=0$. This modification
leads to the following variant of the partition function on the cone:
\begin{eqnarray}
\log Z_{\delta}\left(k\right) & = & \frac{1}{2}\int_{\varepsilon^{2}}^{\infty}\frac{dt}{t}K_{\delta}\left(t\right)\,,\,\mbox{with }K_{\delta}\left(t\right)\equiv\mbox{Tr}\left[e^{-t\left(S^{\left(2\right)}+{\cal R}_{k}\right)}\right]\,.\label{eq:one-loop-proper-time-frge-30}
\end{eqnarray}
A simple way of analyzing (\ref{eq:one-loop-proper-time-frge-30})
is to exploit that the precise form of ${\cal R}_{k}$ is largely arbitrary.
In any case ${\cal R}_{k}$ will leave the contribution to the trace coming from the high momentum modes
untouched, while that of the low momentum modes receives an additional factor
$\rho\left(t\right)\sim e^{-k^{2}t}$, or a qualitatively similar
one, which then suppresses the integrand of the $t$-integral at large
$t\gtrsim 1/k^{2}$. Hence, rather than choosing a specific ${\cal R}_{k}$
and computing the resulting $\rho\left(t\right)$, we may equally well
select right away a suitable function $\rho\left(t\right)$ with the correct
properties, $\rho\left(t\gtrsim 1/k^{2}\right)\approx0$ and $\rho\left(t\lesssim 1/k^{2}\right)\approx1$.
The simplest choice is the step function $\rho\left(t\right)=\theta\left(k^{-2}-t\right)$
which, of course, amounts to a version of the Schwinger's proper time
regularization \cite{Schwinger:1951nm,Dittrich:1985yb}. Applied to (\ref{eq:one-loop-proper-time-frge-30})
it yields
\begin{eqnarray}
\log Z_{\delta}\left(k\right) & = & \frac{1}{2}\int_{\varepsilon^{2}}^{k^{-2}}\frac{dt}{t}K_{\delta}\left(t\right)\,,\label{eq:one-loop-EA-with-deficit-angle-and-IR-cutoff-31}
\end{eqnarray}
with the same kernel $K_{\delta}\left(t\right)$ as in (\ref{eq:one-loop-EA-with-deficit-angle-20}).

By taking the $k$-derivative of (\ref{eq:one-loop-EA-with-deficit-angle-and-IR-cutoff-31})
we can get rid of the UV cutoff $\varepsilon$ at this point: $k\partial_{k}\log Z_{\delta}\left(k\right)=-K_{\delta}\left(k^{-2}\right)$.
Associating a scale dependent entropy ${\cal S}\left(k\right)$ to
$Z_{\delta}\left(k\right)$ via (\ref{eq:S-via-Z-with-def-angle-delta-5}),
we obtain
\begin{eqnarray*}
k\partial_{k}{\cal S}\left(k\right) & = & -\lim_{\delta\rightarrow0}\left[1+2\pi\frac{d}{d\delta}\right]K_{\delta}\left(k^{-2}\right)
\end{eqnarray*}
which evaluates to the following simple RG equation for the, now scale dependent,
entanglement entropy:
\begin{eqnarray}
k\partial_{k}{\cal S}\left(k\right) & = & -\frac{\nu}{24\pi}A\left[\bar{g}\right]k^{2}\,.\label{eq:running-EE-1-50}
\end{eqnarray}
Here we wrote $A\equiv A\left[\bar{g}\right]$ to emphasize the fact
that $A$ is a \emph{proper area} with respect to a classical, externally
prescribed metric, $\bar{g}_{\alpha\beta}$. 

By adopting the discussion
in \cite{Solodukhin:2011gn} it is easy to see that (\ref{eq:running-EE-1-50})
holds not only in flat space but also yields the leading scale dependence
on curved classical spacetimes with any metric $\bar{g}_{\alpha\beta}$. 
Furthermore, equation (\ref{eq:running-EE-1-50})
is equivalent to the RG equation discussed in \cite{Jacobson:2012ek} which
employs a more general cutoff.\\
{\bf (C)} At this point we want to emphasize that in the EAA framework one usually
regards the RG equations, requiring no UV cutoff, as having a more
fundamental status than the functional integral from which they are
derived in a formal way only.\footnote{That is, in presence of a UV regulator.}
In particular this is the stance taken in the Asymptotic Safety program.
This concerns not only the RG equations for the running couplings which parametrize the action functional
$\Gamma_{k}$ itself, but also the RG equations for the {\it co-evolving running parameters}
appearing, for example, in composite operators or observables that do
not correspond to terms in $\Gamma_{k}$ \cite{Pagani:2016dof,Pagani:2016pad,Pagani:2017tdr}. 
In this sense, the entanglement entropy ${\cal S}\left(k\right)$ is an example of
the latter case. Conceptually speaking, it is an ``observable''
that has a scale dependence in its own right, its RG running depends
on the EAA, $\Gamma_{k}$, at least in sufficiently complex truncations. 

Like the EAA itself, the co-evolving quantities, too, are defined in the ``continuum limit'' on
the basis of their RG flow. Hence the UV renormalization problem
translates into the task of finding complete, i.e.~fully extended,
solutions (trajectories) to {\it all} RG equations, those of the co-evolving quantities included \cite{RSBook}.

Let us illustrate this shifted viewpoint by the example of ${\cal S}\left(k\right)$.
While, conceptually speaking, we consider ${\cal S}\left(k\right)$ a co-evolving
quantity with respect to some trajectory of $\Gamma_{k}$, equation (\ref{eq:running-EE-1-50})
happens to be simple enough to require no input from $\Gamma_{k}$
to be integrated:
\begin{eqnarray}
{\cal S}\left(k_{2}\right)-{\cal S}\left(k_{1}\right) & = & -\frac{\nu}{48\pi}A\left[\bar{g}\right]\left(k_{2}^{2}-k_{1}^{2}\right)\,.\label{eq:integrated-flow-EE-1-51}
\end{eqnarray}
This difference of two entropies is the contribution of the field
modes with momenta in the interval $\left[k_{1},k_{2}\right]$. We
are particularly interested in the limits \emph{$k_{1}\rightarrow0$
}and $k_{2}\rightarrow\infty$.
The first limit is unproblematic, yielding
\begin{eqnarray}
{\cal S}\left(0\right)-{\cal S}\left(k_{2}\right) & = & \frac{\nu}{48\pi}A\left[\bar{g}\right]k_{2}^{2}\,.\label{eq:integrated-flow-EE-1-from-zero-52}
\end{eqnarray}
Obviously $k_{2}$ plays the role of the UV cutoff here. In the jargon
of standard field theory one would refer to ${\cal S}\left(k_{2}\right)$
as the ``bare'', and to ${\cal S}\left(0\right)$ as the ``renormalized''
or ``physical'' quantity. From the EAA perspective, equation (\ref{eq:integrated-flow-EE-1-from-zero-52})
corresponds to a finite segment of the RG trajectory, $\left\{ \Gamma_{k},\,k\in\left[0,k_{2}\right]\right\} $,
whose lower endpoint $\Gamma_{0}=\Gamma$ equals the standard effective
action (with a UV cutoff), having an associated entropy ${\cal S}\left(0\right)$.

It remains to take the second limit, $k_{2}\rightarrow \infty$, in which
$\Gamma_{k\rightarrow\infty}$ is known to approach the classical
(bare) action, $S$, essentially. The natural value of the associated
entropy is ${\cal S}\left(k_{2}\rightarrow\infty\right)=0$ since
$\Gamma_{k\rightarrow\infty}$ defines the limiting case of the effective
field theory with no quantum fluctuations integrated out yet. Now,
ideally, we would let $k_{2}\rightarrow\infty$ in equation (\ref{eq:integrated-flow-EE-1-from-zero-52}),
keeping ${\cal S}\left(k_{2}\right)=0$ fixed, and thereby obtain
a finite physical value for the entropy, ${\cal S}\left(0\right)$.
But clearly this is thwarted by the $k_{2}^{2}$-dependence on the
RHS of (\ref{eq:integrated-flow-EE-1-from-zero-52}) which causes
${\cal S}\left(0\right)\propto k_{2}^{2}$ to diverge. In this manner
we re-discover the quadratic divergence of the entanglement entropy
in the framework of the EAA. It is signalled by the non-existence
of an RG trajectory that extends to all $k\in\left[0,\infty\right)$.

Next let us see how the situation changes in quantum gravity.

\section{Entanglement entropy in QEG \label{sec:Entanglement-entropy-in-QEG}}

Up to now we considered matter fields in a prescribed classical background
spacetime. Now we go on to Quantum Einstein Gravity (QEG) as defined
by a complete, asymptotically free RG trajectory $\Gamma_{k}\left[h_{\alpha\beta},\Phi;\bar{g}_{\alpha\beta}\right]$,
$k\in\left[0,\infty\right)$. As usual, $h_{\alpha\beta}$ and $\bar{g}_{\alpha\beta}$
denote the metric fluctuation and the background metric, respectively.\footnote{We suppress the Faddeev-Popov ghosts here.}
While our arguments are general, in explicit calculations we will
employ the single-metric Einstein-Hilbert truncation coupled to the
matter fields $\Phi$; the only running gravitational couplings are
the Newton's constant $G\left(k\right)$ and the cosmological constant
$\Lambda\left(k\right)$ then \cite{R98,Reuter:2001ag}. It is assumed that
the matter fields combined in $\Phi$ are such that they do not destroy
the Asymptotic Safety of pure gravity \cite{Percacci:2002ie}.

We may also assume that the \emph{reconstruction problem} \cite{Manrique:2008zw}
has been solved within the truncation considered. As a result, we have a regularized
functional integral at our disposal,
\[
\int{\cal D}_{\Lambda}\hat{h}\,{\cal D}_{\Lambda}\hat{\Phi}\,e^{-S\left[\hat{h},\hat{\Phi};\bar{g}\right]}\,,
\]
which approaches a well defined limit when its UV regulator is removed
($\Lambda\rightarrow\infty$) and which reproduces the asymptotically safe
RG trajectory. (See \cite{Manrique:2008zw} for a detailed discussion.) \\
{\bf (A)} What is the meaning of the calculation in section \ref{sec:Entanglement-entropy-on-rigid-background}
in the Asymptotic Safety context, if any? First of all, as it stands
the result for the entanglement entropy refers to a free field. So
let us assume that among the matter fields $\Phi$ there is at least
one that appears quadratically in the fixed point action\footnote{Presumably, this is not very restrictive
\cite{Percacci:2015wwa,Christiansen:2017cxa}.},
and let us compute its contribution to the entanglement entropy. 

In the gravitational EAA approach, Background Independence is established
by studying the dynamics of the metric fluctuation $h_{\alpha\beta}=g_{\alpha\beta}-\bar{g}_{\alpha\beta}$
and matter fields \emph{on all backgrounds simultaneously}, i.e.~$\bar{g}_{\alpha\beta}$
should be left completely arbitrary in the calculation of $\Gamma_{k}$
and the concomitant running quantities.

In this spirit, we now interpret equation (\ref{eq:running-EE-1-50})
as the result of a calculation in an arbitrary but fixed, classical
background metric, $\bar{g}_{\alpha\beta}$. So at this point $k\partial_{k}{\cal S}\left(k\right)$
should be understood as a {\it functional of $\bar{g}_{\alpha\beta}$}.\\
{\bf (B)} Quantum gravity, and specifically QEG, differs most fundamentally
from any standard quantum field theory in that it must dynamically
generate the spacetime geometry in which all other physics is going
to take place then. In particular the theory should be able to distinguish
physically realistic, stable states $|\psi\rangle$ from unstable
or impossible ones that would never be seen in Nature. The EAA encodes
information about physically acceptable states via the metric expectation
value it gives rise to, $\langle\psi|\hat{g}_{\alpha\beta}|\psi\rangle=g_{\alpha\beta}$.

In the background field formalism, knowing $\Gamma_{k}$, we can search
for \emph{self-consistent background metrics }$\bar{g}_{\alpha\beta}=\left(\bar{g}_{k}^{{\rm sc}}\right)_{\alpha\beta}$.
By definition, when the $h_{\alpha\beta}$ fluctuations (and the matter
fields) are quantized in a self-consistent background, $\hat{h}_{\alpha\beta}$
has vanishing expectation value, $\langle\hat{h}_{\alpha\beta}\rangle\equiv h_{\alpha\beta}=g_{\alpha\beta}-\bar{g}_{\alpha\beta}=0$,
and so $g_{\alpha\beta}=\bar{g}_{\alpha\beta}$. In these special
backgrounds the quantum fluctuations are particularly tame, and we
may regard $g_{\alpha\beta}=\bar{g}_{\alpha\beta}=\left(\bar{g}_{k}^{{\rm sc}}\right)_{\alpha\beta}$
as the expectation value of the metric operator in a physically realistic
state.\footnote{Note that deciding for a self-consistent background is more special
than merely ``going on-shell''. When split symmetry is broken, the
two notions are inequivalent since a non-zero $h_{\alpha\beta}=\langle\hat{h}_{\alpha\beta}\rangle$
cannot straightforwardly be absorbed into the background metric. Recall
also \cite{Becker:2014pea} that the general effective field equation for
configurations $h_{\alpha\beta}\neq0$ is more complicated than the
tadpole equation (\ref{eq:tadpole-defining-selfconsistent-background-59})
as it contains an additional term $\propto{\cal R}_{k}h_{\alpha\beta}$,
which would affect the argument below.} 

Self-consistent backgrounds are found by solving the tadpole equation \cite{Becker:2014pea}:
\begin{eqnarray}
\frac{\delta}{\delta h_{\alpha\beta}\left(x\right)}\Gamma_{k}\left[h;\bar{g}\right]\Bigr|_{h=0,\, \bar{g}=\bar{g}_{k}^{{\rm sc}}} 
& = & 0\,.\label{eq:tadpole-defining-selfconsistent-background-59}
\end{eqnarray}

Moreover, thanks to the tadpole equation (\ref{eq:tadpole-defining-selfconsistent-background-59}), 
the self-consistent background can also be employed to compute the partition function of the system.
A detail discussion regarding the properties of such a partition function can be found in \cite{Becker:2014pea}.
The so computed partition function is then a functional of the self-consistent background: $Z\left[\bar{g}_{k}^{{\rm sc}}\right]$.

According to the discussion of section \ref{sec:Entanglement-entropy-on-rigid-background}, in order to 
compute the entanglement entropy via the replica trick, one must introduce a deficit angle in the geometry of the
system and remove the singularity at the end of the calculation.
Namely, one must evaluate the quantity $Z\left[\bar{g}_{k,\delta}^{{\rm sc}}\right]$,
where $\delta$ is the deficit angle.
Note that in general there is no reason for $Z\left[\bar{g}_{k,\delta}^{{\rm sc}}\right]$ to be determined directly
by $Z\left[\bar{g}_{k}^{{\rm sc}}\right]$ since the introduction of the deficit angle changes the topology
of the spacetime and a new calculation is required.\footnote{
If the EAA was computed keeping track also of the topology dependence,
then it may be possible to evaluate the entanglement entropy directly from the EAA itself.}

Focusing on the Einstein-Hilbert truncation now, the tadpole equation
happens to have the same structure as the classical Einstein equation:
\begin{eqnarray}
R_{\,\,\:\nu}^{\mu}\left(\bar{g}_{k}^{{\rm sc}}\right)-\frac{1}{2}\delta_{\nu}^{\mu}R\left(\bar{g}_{k}^{{\rm sc}}\right)+\Lambda\left(k\right)\delta_{\nu}^{\mu} & = & 0\,.\label{eq:tadpole-for-EH-truncation-60}
\end{eqnarray}
Since under rescalings of the metric the Ricci tensor behaves as $R_{\,\,\:\nu}^{\mu}\left(c^{2}\bar{g}_{k}^{{\rm sc}}\right)=c^{-2}R_{\,\,\:\nu}^{\mu}\left(\bar{g}_{k}^{{\rm sc}}\right)$,
it follows that solutions to (\ref{eq:tadpole-for-EH-truncation-60})
respond to changes of the cosmological constant in such a way that $\Lambda\left(k\right)\left(\bar{g}_{k}^{{\rm sc}}\right)_{\alpha\beta}=\,$const.
It proves convenient to introduce an arbitrary normalization scale
$\mu$ in order to write this relation as $\Lambda\left(k\right)\left(\bar{g}_{k}^{{\rm sc}}\right)_{\alpha\beta}=\Lambda\left(\mu\right)\left(\bar{g}_{\mu}^{{\rm sc}}\right)_{\alpha\beta}$,
or as
\begin{eqnarray}
\left(\bar{g}_{k}^{{\rm sc}}\right)_{\alpha\beta} & = & \frac{\Lambda\left(\mu\right)}{\Lambda\left(k\right)}\left(\bar{g}_{\mu}^{{\rm sc}}\right)_{\alpha\beta}\,=\,\frac{\mu^{2}\lambda\left(\mu\right)}{k^{2}\lambda\left(k\right)}\left(\bar{g}_{\mu}^{{\rm sc}}\right)_{\alpha\beta}\,.\label{eq:g-selfconsisten-via-lambdaS-and-g_mu-61}
\end{eqnarray}
In the second equality we inserted the dimensionless cosmological
constant $\lambda\left(k\right)=\Lambda\left(k\right)/k^{2}$, and
correspondingly for $k=\mu$.

Likewise we redefine the field variables by writing them as dimensionless
multiples of the cutoff, or appropriate powers thereof. The dimensionless
metric coefficients are then $\tilde{g}_{\alpha\beta}\equiv k^{2}g_{\alpha\beta}$
and $\tilde{\bar{g}}_{\alpha\beta}\equiv k^{2}\bar{g}_{\alpha\beta}$,
and so, from (\ref{eq:g-selfconsisten-via-lambdaS-and-g_mu-61}):
\begin{eqnarray}
\widetilde{\left(\bar{g}_{k}^{{\rm sc}}\right)}_{\alpha\beta} & = & \frac{\lambda\left(\mu\right)}{\lambda\left(k\right)}\widetilde{\left(\bar{g}_{\mu}^{{\rm sc}}\right)}_{\alpha\beta}\,.\label{eq:dimless-g-selfconsistent-via-lambdaS-62}
\end{eqnarray}

Let us recall that in the case of an asymptotically safe UV limit
it is the \emph{dimensionless }couplings that assume fixed point values.
For instance, $\lambda\left(k\right)$ approaches a finite number
$\lim_{k\rightarrow\infty}\lambda\left(k\right)=\lambda_{*}$. Accordingly,
it is the dimensionless form of the tadpole equation that continues
to be meaningful in the limit of $k\rightarrow\infty$, admitting
a finite solution
\begin{eqnarray}
\widetilde{\left(\bar{g}_{*}^{{\rm sc}}\right)}_{\alpha\beta} & = & \lim_{k\rightarrow\infty}\widetilde{\left(\bar{g}_{k}^{{\rm sc}}\right)}_{\alpha\beta}\,=\,\frac{1}{\lambda_{*}}\lambda\left(\mu\right)\widetilde{\left(\bar{g}_{\mu}^{{\rm sc}}\right)}_{\alpha\beta}\,.\label{eq:dimless-g-selfconsistent-fixed-point-63}
\end{eqnarray}
\\
{\bf (C)} After these preparations we return to the entanglement entropy and
reconsider the calculation of section \ref{sec:Entanglement-entropy-on-rigid-background}
within QEG. In order to obtain the corresponding entropy ${\cal S}_{{\rm QEG}}\left(k\right)$
we choose $\bar{g}_{\alpha\beta}$ in the final result for a rigid
background, equation (\ref{eq:integrated-flow-EE-1-from-zero-52}),
to be a self-consistent one for the corresponding scale, $\left(\bar{g}_{k}^{{\rm sc}}\right)_{\alpha\beta}$.
In this manner we obtain the entanglement entropy related to a system
of fields inhabiting a spacetime which is indeed physically realizable,
or, at the very least, is much closer to a realizable one than it would be on a generic
background. Clearly this is a necessary prerequisite if the entropy
computed is to have a physical meaning, and hence a reason to be finite.

Thus we obtain from equation (\ref{eq:integrated-flow-EE-1-from-zero-52}),
writing $k=k_{2}$ from now on, 
\begin{eqnarray}
{\cal S}_{{\rm QEG}}\left(0\right)-{\cal S}_{{\rm QEG}}\left(k\right) & = & \frac{\nu}{48\pi}A\left[\bar{g}_{k}^{{\rm sc}}\right]k^{2}\nonumber \\
 & = & \frac{\nu}{48\pi}A\left[k^{2}\bar{g}_{k}^{{\rm sc}}\right]\,.\label{eq:integrated-flow-EE-QEG-selfconsistent-g-65}
\end{eqnarray}
Here we also exploited the fact that the area scales as $A\left[c^{2}g_{\alpha\beta}\right]=c^{2}A\left[g_{\alpha\beta}\right]$.
As a result, the entropy difference (\ref{eq:integrated-flow-EE-QEG-selfconsistent-g-65})
depends on the scale $k$ only via the \emph{dimensionless }self-consistent
metric, that is $\widetilde{\left(\bar{g}_{k}^{{\rm sc}}\right)}$:
\begin{eqnarray}
{\cal S}_{{\rm QEG}}\left(0\right)-{\cal S}_{{\rm QEG}}\left(k\right) & = & \frac{\nu}{48\pi}A\left[\widetilde{\left(\bar{g}_{k}^{{\rm sc}}\right)}\right]\,.\label{eq:integrated-flow-EE-QEG-selfconst-dimless-g-66}
\end{eqnarray}
Remarkably enough, when we let $k\rightarrow\infty$ the quantity
(\ref{eq:integrated-flow-EE-QEG-selfconst-dimless-g-66}) approaches
a well defined limit ${\cal S}_{{\rm QEG}}\left(0\right)-{\cal S}_{{\rm QEG}}\left(\infty \right)\equiv\Delta{\cal S}_{{\rm QEG}}$:
\begin{eqnarray}
\Delta{\cal S}_{{\rm QEG}} & = & \frac{\nu}{48\pi}A\left[\widetilde{\left(\bar{g}_{*}^{{\rm sc}}\right)}\right]\,.\label{eq:integrated-flow-EE-QEG-selfconst-dimless-g-FP-67}
\end{eqnarray}
This perfectly finite result for the entanglement entropy in QEG is
our main result.
\\
{\bf (D)} Using (\ref{eq:dimless-g-selfconsistent-fixed-point-63}) we may rewrite (\ref{eq:integrated-flow-EE-QEG-selfconst-dimless-g-FP-67})
in the more practically applicable forms
\begin{eqnarray}
\Delta{\cal S}_{{\rm QEG}} & = & \frac{\nu}{48\pi}\frac{\lambda\left(\mu\right)}{\lambda_{*}}\, \mu^{2}\, {\cal A}\left(\mu\right)
\label{eq:integrated-flow-EE-QEG-dimless-area-g-FP-68}\\
 & = & \frac{\nu}{48\pi}\frac{\Lambda\left(\mu\right)}{\lambda_{*}}\, {\cal A}\left(\mu\right)\,,\nonumber 
\end{eqnarray}
where ${\cal A}\left(\mu\right)\equiv A\left[\bar{g}_{\mu}^{{\rm sc}}\right]$
is the dimensionful proper area measured with the background metric
at the normalization point $\mu$. 

It needs to be emphasized though
that \emph{the entanglement entropy is independent of the normalization
scale }$\mu$. In (\ref{eq:g-selfconsisten-via-lambdaS-and-g_mu-61})
we introduced $\mu$ in such a way that the product $\Lambda\left(\mu\right)\left(\bar{g}_{\mu}^{{\rm sc}}\right)_{\alpha\beta}=\lambda\left(\mu\right)\mu^{2}\left(\bar{g}_{\mu}^{{\rm sc}}\right)_{\alpha\beta}$
stays constant when $\mu$ is changed, hence $\Lambda\left(\mu\right){\cal A}\left(\mu\right)=\lambda\left(\mu\right)\mu^{2}{\cal A}\left(\mu\right)$
and therefore the entropy are $\mu$-independent:
\begin{eqnarray}
\mu\frac{d}{d\mu}\left\{ \lambda\left(\mu\right)\mu^{2}{\cal A}\left(\mu\right)\right\}  & = & 0\,.\label{eq:CS-for-EE-69}
\end{eqnarray}
Equation (\ref{eq:CS-for-EE-69}) may be seen as a simple example
of a Callan-Symanzik equation.\\
{\bf (E)} Let us consider the following thought experiment to determine the
entanglement entropy related to a given surface $\Sigma$. In order to measure
the area of $\Sigma$ we must choose a specific ``yard stick'' (or
``microscope''); it is characterized by a certain minimal length
which it is able to resolve, $\ell$. The actual measurement consists in
using this yard stick to partition $\Sigma$ in little squares of
side length $\ell$, and counting the resulting ``pixels''; let ${\cal N}\left(\ell \right)$
denote their total number.

We may assume that the best possible effective field theory description
of this measuring procedure is obtained from that EAA which has its scale $k$,
or in the present case $\mu$, adapted to the scale of the experiment,
$\mu\approx \ell^{-1}$. Hence the measurement is described as taking
place in the classical spacetime geometry with $g_{\mu}^{{\rm sc}}\Bigr|_{\mu=\ell^{-1}}$.
Recall also \cite{Reuter:2005bb} that the length scale $k^{-1}$ pertaining
to $\Gamma_{k}\left[h;\bar{g}\right]$ is a proper length with respect
to its second argument, $\bar{g}_{\alpha\beta}$. As a consequence,
we can say that the little squares we counted have the proper area
$\ell^{2}=\mu^{-2}$ with respect to the optimum self-consistent background
metric $g_{\ell^{-1}}^{{\rm sc}}$. Hence the result of counting pixels,
${\cal N}\left(\ell\right)$, has the following interpretation within
the effective field theory:
\begin{eqnarray}
{\cal N}\left(\ell \right) & = & \frac{A\left[\bar{g}_{\ell^{-1}}^{{\rm sc}}\right]}{\ell^{2}}\,\equiv\,\mu^{2}{\cal A}\left(\mu\right)\Bigr|_{\mu=\ell^{-1}}\,.\label{eq:cal-N-via-dimless-A-80}
\end{eqnarray}
If we accept this interpretation, along with equation (\ref{eq:integrated-flow-EE-QEG-dimless-area-g-FP-68}),
we can deduce the desired entropy from our pixel count:
\begin{eqnarray}
\Delta{\cal S}_{{\rm QEG}} & = & \frac{\nu}{48\pi\lambda_{*}}\lambda\left(1/\ell\right){\cal N}\left(\ell \right)\,.\label{eq:Delta-S-QEG-via-calN-81}
\end{eqnarray}
Again, ${\cal N}\left(\ell \right)$ will not be independent of $\ell$ in
general, but the product $\lambda\left(1/\ell \right){\cal N}\left(\ell \right)$,
and hence the entropy, are $\ell$-independent.

This has an important consequence: If the RG trajectory, and in particular
the function $\lambda\left(k\right)$ are known, we can determine the
entropy on the basis of the formula (\ref{eq:integrated-flow-EE-QEG-dimless-area-g-FP-68})
by performing the experiment \emph{on any scale we like}. This may lead
to different numbers of pixels, but the resulting entropy is always
the same \emph{provided the running of the cosmological constant is
taken into account properly}.

For example, we could decrease $\ell$ to the point that $\mu=1/\ell \rightarrow\infty$
enters the scaling regime of the UV fixed point so that $\lambda\left(1/\ell \right)\rightarrow\lambda_{*}$.
This limit gives rise to the following representation of the entropy:
\begin{eqnarray}
\Delta{\cal S}_{{\rm QEG}} & = & \frac{\nu}{48\pi}{\cal N}_{*}\,,\,\mbox{where }{\cal N}_{*}=\lim_{\ell \rightarrow0}{\cal N}\left(\ell \right)\,.\label{eq:Delta-S_QEG-via-Npixel-FP-82}
\end{eqnarray}
As soon as the RG trajectory reaches the fixed point regime, $\lambda\left(\mu\right)$
stops running. Hence, by (\ref{eq:CS-for-EE-69}), the area scales
as ${\cal A}\left(\mu\right)\propto1/\mu^{2}$ so that ${\cal N}\left(\ell \right)$
becomes independent of $\ell$; it no longer increases when $\ell$ is decreased
even further.

If the EAA follows a type IIIa trajectory \cite{Reuter:2001ag} which has
a long classical regime in the infrared we can use the constant value
of the Newton's constant, $G_{{\rm class}}$, in order to define Planck
units, $\ell _{{\rm Pl}}\equiv1/m_{{\rm Pl}}\equiv\sqrt{G_{{\rm class}}}$.
Picking $\mu=m_{{\rm Pl}}$ leads to a representation of the entanglement
entropy that comes close to the Bekenstein-Hawking formula:
\begin{eqnarray}
\Delta{\cal S}_{{\rm QEG}} & = & \frac{\nu}{12\pi}\frac{\lambda\left(m_{{\rm Pl}}\right)}{\lambda_{*}}\frac{{\cal A}\left(m_{{\rm Pl}}\right)}{4G_{{\rm Pl}}}\,.\label{eq:Delta-S_QEG-Bekenstein-like-90}
\end{eqnarray}
Note however that the prefactor in equation (\ref{eq:Delta-S_QEG-Bekenstein-like-90}),
while in fact generically of order unity, depends on the matter contents
both via $\nu$ and the trajectory, i.e., via the ratio $\lambda\left(m_{{\rm Pl}}\right)/\lambda_{*}$.

Finally, let us emphasize that, even if we considered the Einstein-Hilbert truncation for the gravitation EAA,
the scaling behaviour of the dimensionless self-consistent metric and of the associated entropy (\ref{eq:integrated-flow-EE-QEG-selfconst-dimless-g-66})
is an exact consequence of the Asymptotic Safety scenario as such.
It follows that a finite entanglement entropy is achieved also in the case of more refined gravitational EAA truncations.
In particular, such extended truncations include
higher curvature truncations \cite{Lauscher:2002sq,Reuter:2002kd,
Codello:2007bd,Benedetti:2009rx,Benedetti:2010nr,Rechenberger:2012pm,
  Ohta:2013uca,Benedetti:2013jk,Falls:2014tra,Eichhorn:2015bna,
  Ohta:2015efa,Falls:2016wsa,Falls:2016msz,Gies:2016con},
$f(R)$ and infinite dimensional truncations 
\cite{Reuter:2008qx,Benedetti:2012dx,Demmel:2012ub,Dietz:2012ic,Bridle:2013sra,Dietz:2013sba,Demmel:2014sga,
Demmel:2014hla,Demmel:2015oqa,Ohta:2015fcu,Labus:2016lkh,Dietz:2016gzg,Christiansen:2017bsy,Knorr:2017mhu,Falls:2017lst,Alkofer:2018fxj},
bimetric truncations
 \cite{Manrique:2009uh,Manrique:2010mq,Manrique:2010am,Christiansen:2012rx,Codello:2013fpa,Christiansen:2014raa,
 Becker:2014qya,Christiansen:2015rva,Knorr:2017fus}, 
truncations for extended theories of gravity 
\cite{Daum:2010qt,Daum:2013fu,Pagani:2013fca,Pagani:2015ema,Reuter:2015rta},
truncations on foliated spacetimes
\cite{Manrique:2011jc,Rechenberger:2012dt,Biemans:2016rvp,Biemans:2017zca,Houthoff:2017oam},
and truncations with different kinds of matter content 
\cite{Dou:1997fg,Percacci:2002ie,Narain:2009fy,Daum:2010bc,
  Folkerts:2011jz,Harst:2011zx,Eichhorn:2011pc,
  Eichhorn:2012va,Dona:2012am,Dona:2013qba,Labus:2015ska,Oda:2015sma,Meibohm:2015twa,Dona:2015tnf,
  Meibohm:2016mkp,Eichhorn:2016esv,Eichhorn:2016vvy,
  Christiansen:2017gtg,Eichhorn:2017ylw,Eichhorn:2017lry}.
In particular it would be interesting to compute the entanglement entropy for those RG trajectories
that are compatible with unitarity \cite{Becker:2017tcx,Arici:2017whq}.

\section{Summary}

When defined on a rigid classical spacetime geometry, quantized matter
fields are known to give rise to an entanglement entropy which is
proportional to the area of the entangling surface, with a factor
of proportionality which is quadratically divergent though. 
In this paper we employed instead a background independent approach to quantum gravity
and regarded
the entanglement entropy as a scale dependent quantity which RG-evolves
in parallel with the Effective Average Action. The latter controls
the geometry of spacetime at the mean field level, among other things,
and in particular it determines the self-consistent background geometries for each
scale. 
The leading term of the entanglement entropy in those
geometries turned out to be perfectly finite. The cutoff dependence
of the entropy is precisely cancelled by the RG running of the metric in the
infinite cutoff limit. 

While, for illustrative purposes, we considered
the Einstein-Hilbert truncation here, the finiteness of the entropy
is a direct consequence of Asymptotic Safety as such and it applies also 
to more refined truncations schemes.

All that is required is the scaling behaviour of the metric corresponding to a non-Gaussian
UV fixed point. Hence the finiteness of the leading entropy term is obtained analogously
in $d$ spacetime dimensions for surfaces $\Sigma$ of co-dimension two.

From the perspective of Asymptotic Safety, the notorious quadratic
UV divergence seems to occur because one is asking an unphysical
question, and tries to compute a quantity that never could be measured in Nature,
not even in principle. The divergence disappears as soon as we admit that,
at asymptotically high scales, spacetime is actually fractal like \cite{Lauscher:2005qz}, and
carries a metric which strongly depends on the ``length of the yard
stick'' that is used to probe the spacetime.


\clearpage



\end{spacing}


\end{document}